\documentclass[authoryear,preprint,review,12pt]{elsarticle}

\usepackage{amssymb}
\usepackage{amsmath}
\usepackage{booktabs}
\usepackage{multirow}
\usepackage{graphicx} 
\usepackage{comment}
\usepackage{lineno}
\usepackage{tabularx} 

\begin{document}


\begin{frontmatter}



\title{From Narrative to Action: A Hierarchical LLM-Agent Framework for Human Mobility Generation} 

\author[a,b]{Qiumeng Li\corref{cor1}}
\cortext[cor1]{Corresponding author.}
\ead{qiumengli@hkust-gz.edu.cn}
\author[a]{Chunhou Ji}
\author[a]{Xinyue Liu}

\address[a]{Thrust of Urban Governance and Design, Society Hub, Hong Kong University of Science and Technology (Guangzhou), Guangzhou, Guangdong, China}
\address[b]{Thrust of Intelligent Transportation, System Hub, Hong Kong University of Science and Technology (Guangzhou), Guangzhou, Guangdong, China}

\begin{abstract}
Understanding and replicating human mobility requires not only spatial-temporal accuracy but also an awareness of the cognitive hierarchy underlying real-world travel decisions. Traditional agent-based or deep learning models can reproduce statistical patterns of movement but fail to capture the semantic coherence and causal logic of human behavior. Large language models (LLMs) show potential, but struggle to balance creative reasoning with strict structural compliance. This study proposes a Hierarchical LLM-Agent Framework, termed Narrative-to-Action, that integrates high-level narrative reasoning, mid-level reflective planning, and low-level behavioral execution within a unified cognitive hierarchy. At the macro level, one agent is employed as a "creative writer" to produce diary-style narratives rich in motivation and context, then uses another agent as a "structural parser" to convert narratives into machine-readable plans. A dynamic execution module further grounds agents in geographic environments and enables adaptive behavioral adjustments guided by a novel occupation-aware metric, Mobility Entropy by Occupation (MEO), which captures heterogeneous schedule flexibility across different occupational personalities. At the micro level, the agent executes concrete actions-selecting locations, transportation modes, and time intervals-through interaction with an environmental simulation. By embedding this multi-layer cognitive process, the framework produces not only synthetic trajectories that align closely with real-world patterns but also interpretable representations of human decision logic. This research advances synthetic mobility generation from a data-driven paradigm to a cognition-driven simulation, providing a scalable pathway for understanding, predicting, and synthesizing complex urban mobility behaviors through hierarchical LLM agents.
\end{abstract}



\end{frontmatter}



\section{Introduction}

Human mobility patterns are fundamental drivers of urban dynamics(\cite{fuller2017analysis}). Large-scale, high-precision individual activity trajectory data, which record information about when, where, and what activities people engage in, have become a key enabler for next-generation smart city applications(\cite{boeing2022urban, yang2023identifying}). From optimizing public transportation networks and strategically planning commercial and public facilities to simulating traffic flows and evaluating emergency evacuation strategies, such trajectory data offer an unprecedented microscopic perspective for understanding and improving urban systems(\cite{yang2023identifying}). However, directly obtaining and utilizing real-world trajectory data face two major challenges. In terms of privacy, GPS and mobile positioning data capture some of the most sensitive personal information of citizens, and their use is strictly regulated by data protection laws worldwide (\cite{fiore2019privacy, sookhak2018security})(e.g., GDPR(\cite{voigt2017eu})). In terms of cost and sparsity, the high cost of data collection and insufficient sample coverage make it nearly impossible to construct comprehensive real-world datasets.

To address these challenges, the generation of high-quality synthetic human trajetory data has emerged as a promising alternative(\cite{kapp2023generative, uugurel2024learning}). Ideally, synthetic data should preserve the statistical properties of real-world data while effectively mitigating privacy risks (\cite{fiore2019privacy, kapp2023generative, miranda2023sok}), thereby providing secure and reliable support for downstream research. 
In prior studies, researchers have proposed various generation approaches. Early rule-based or statistical models, such as those grounded in time geography or Markov chains, could construct basic activity sequences(\cite{lu2013approaching, song2010modelling}). However, their inherent determinism and simplified assumptions often resulted in highly uniform and overly rigid trajectories, failing to capture the flexibility and individuality inherent in human behavior(\cite{lu2013approaching}). Subsequent deep learning approaches, such as recurrent neural networks (RNNs) and generative adversarial networks (GANs), have demonstrated improvements in learning data distributions(\cite{berke2022generating, mauro2022generating}). Nevertheless, as these models essentially fit spatiotemporal coordinate sequences, they often generate trajectories that lack semantic coherence or spatial plausibility, due to their limited capacity for deep understanding of real-world logic(\cite{kapp2023generative, merhi2024synthetic}).
In recent years, the rapid progress of LLMs has opened new possibilities for human activity and mobility sequence generation tasks(\cite{achiam2023gpt,liu2024deepseek,team2024gemini}). Unlike earlier statistical or deep learning approaches, LLMs are capable of producing semantically rich descriptions of daily activities. Researchers have begun leveraging carefully crafted prompts to directly instruct LLMs to output structured activity plans—often in machine-readable formats for downstream processing(\cite{zhao2023survey}). However, this “single-stage” direct generation approach faces a fundamental dilemma: it forces the model to balance two inherently conflicting objectives—creative reasoning and strict format compliance. As a result, outputs are often suboptimal. In some cases, strict adherence to the predefined format comes at the expense of narrative richness and individuality, yielding logically correct but bland plans. In others, the pursuit of creativity and detail leads to poorly formatted or even logically inconsistent results, failing to capture the coherent “storyline” that underlies human behavior.

Existing LLM-driven mobility studies remain largely flat in structure: they often generate static activity sequences or narrative descriptions without incorporating the hierarchical cognition that underlies real human behavior. To bridge this gap, the paper introduce \textbf{a Hierarchical LLM-Agent Framework, termed Narrative-to-Action}, which operationalizes a multi-layer cognitive architecture for human mobility generation and simulation. The framework includes high-level narrative reasoning, mid-level reflective planning and low-level behavioral execution, allowing agents to transform human-like stories into adaptive, spatially grounded actions. At the macro level, the framework implements a two-stage narrative-driven generation by decoupling creative reasoning from structured output. The process first guides one LLM to produce a diary-style narrative rich in motivations, contexts, and causal logic, and then employs a second LLM to parse the narrative into a machine-readable activity plan with semantic locations. This narrative–parsing design enables each model to focus on a single objective, thereby avoiding unnecessary trade-offs between creativity and structural validity. 

Building on this foundation, the study further extends it with dynamic execution modules that allow agents to continuously adapt their behavior during simulation rather than rigidly following static schedules. Recognizing that individuals with different occupations vary in how strictly they adhere to daily plans, the study refines the adaptive execution stage by introducing Mobility Entropy by Occupation (MEO), a novel measure capturing the inherent variability of daily schedule flexibility across different occupational groups. By embedding occupation-aware flexibility into the reflective decision-making layer, the framework captures differentiated behavioral adaptations across occupational personalities, and personalizes the probability of rethinking, aligning simulated adaptability with real-world socio-economic heterogeneity. Overall, this work advances human mobility simulation from data-driven generation toward \textbf{cognition-driven modeling}, providing both realistic synthetic data and a cognitively interpretable paradigm for understanding the hierarchical structure of human travel decision-making.

\section{Related work}
\subsection{Human Activity Trajectory Generation}
The generation of synthetic human activity trajectories has long been a central topic in urban computing and transportation science. Early efforts relied on statistical models and predefined rules, such as Markov chains, which captured basic constraints and transition probabilities but struggled with long-term dependencies(\cite{chen2025trajectory,  jeong2021variational, jiang2017activity}). Although these methods reproduced certain macroscopic mobility patterns(\cite{song2010limits}), their “memoryless” nature often produced activity sequences that lacked deep coherence(\cite{kulkarni2019examining}). With the advent of deep learning, generative models such as RNNs(\cite{liu2016predicting}, LSTMs(\cite{sun2021joint}), and GANs(\cite{gao2022generative, kulkarni2018generative}) enabled more complex pattern learning and improved realism, yet they still suffered from two key limitations: the lack of semantic and logical constraints(\cite{liao2024deep}), and poor interpretability(\cite{luca2021survey}).

Fundamentally, both statistical and deep learning approaches view trajectory generation as a sequence modeling problem over spatiotemporal points, without explicitly capturing the causal logic and semantic coherence underlying human decision-making. This limitation often leads to activity sequences that appear continuous in time and space but lack behavioral plausibility. Several studies have attempted to address this issue by incorporating temporal variation or semantic constraints into their models(\cite{liao2024deep, wang2024generating, xiong2023trajsgan}). While these efforts represent important progress, they remain limited in capturing the broader narrative and adaptive dynamics of human activity patterns.

Recent advances have introduced LLMs into mobility generation, offering the capacity to produce semantically rich and context-aware activity plans.
TrajLLM(\cite{ju2025trajllm}) employs a modular agent-based framework that integrates persona generation, activity selection, and destination prediction. Geo-Llama(\cite{li2025geo}) fine-tunes LLMs to generate trajectories under both unconstrained and constraint-driven settings while maintaining contextual coherence. In parallel, TrajLLM(\cite{jiawei2024large}) as “virtual citizens” generating heterogeneous personal trajectories, while MobAgent(\cite{li2024more}) and CoPB(\cite{shao2024chain}) leverage socio-demographic profiles and behavioral theory to produce plausible, individualized travel diaries. While these studies mark significant progress, they remain focused on plan generation and seldom incorporate the dynamic execution mechanisms required for simulating adaptive agent behaviors in realistic environments.

\subsection{Agent-Based Simulation and the Paradigm Shift with LLMs}
Agent-based models (ABMs) have long served as a foundational tool in transportation research and urban simulation, representing individuals as decision-making entities and capturing behavioral heterogeneity in complex systems(\cite{w2016multi, krajzewicz2010traffic}). Despite their widespread use and strengths in modeling large-scale dynamics, most traditional ABMs rely on rule-based or discrete choice formulations with fixed utility functions. As a result, activity sequences and mode choices are predetermined rather than emergent, limiting the model’s ability to reflect the adaptive and context-sensitive nature of human decision-making(\cite{wang2025agentic}). To better capture human-like reasoning, researchers have long proposed influential cognitive architectures such as the Belief–Desire–Intention (BDI) model(\cite{rao1995bdi, davidsson2002agent}). The BDI framework assumes that an intelligent agent acts based on its beliefs (representations of the world), desires (goals), and intentions (plans to which it is committed). This model provides a robust theoretical foundation for practical reasoning, yet before the advent of LLMs, implementing it at scale and endowing agents with rich contextual understanding remained highly challenging.

Recent progress in artificial intelligence has introduced the notion of generative agents, which aim to simulate human-like cognition by incorporating memory, planning, and social interaction mechanisms(\cite{park2023generative}). Research shows that when agents are endowed with internal states and narrative continuity, they can exhibit behaviors that appear more natural and adaptive than those produced by traditional rule-based counterparts. Building on this idea, several studies have adapted generative agents for transportation and urban applications. Frameworks such as GATSim (\cite{liu2025gatsim}), AgentSociety(\cite{piao2025agentsociety}), and YulanOneSim(\cite{wang2025yulan}) enrich agents with lifestyle attributes, socio-economic profiles, evolving preferences, and memory, enabling large-scale simulations of more realistic mobility behaviors. By moving beyond rigid rules and incorporating adaptive decision-making processes, these approaches significantly enhance the fidelity of agent-based simulations. However, most existing systems still treat activity plans as exogenous inputs or static schedules, without explicitly addressing the generation of coherent daily plans or integrating plan generation with dynamic execution.

The incorporation of LLMs into social simulation represents a deeper paradigm shift in how agent behavior is specified and how macro-level phenomena emerge. Traditional ABMs define behavior through explicit rules: agents perceive a limited set of environmental variables and act according to conditional statements or static utility functions. While transparent and computationally efficient, this approach confines agents to a predefined action space and prevents them from adapting to unforeseen scenarios. In contrast, LLM-driven agents can be specified at a higher level of abstraction. Rather than coding actions step by step, modelers can define an agent’s role, background, preferences, and objectives in natural language. The LLM then reasons about how to act in novel contexts by combining these identity descriptions with its latent knowledge. Although this does not guarantee correct behavior, it enables agents to respond adaptively and exhibit decision patterns more closely aligned with those observed in human societies(\cite{zheng2025urban}).

\subsection{Computational Narrative and Data Storytelling}
Narrative is a fundamental way in which humans organize experiences, understand causal relationships, and communicate knowledge(\cite{bruner1991narrative}). In artificial intelligence, computational narrative research has explored how systems can generate, understand, and apply stories, showing that a logically structured storyline significantly improves coherence and credibility in generated content(\cite{riedl2010narrative}). However, this perspective has rarely been applied to the modeling of human mobility. Existing trajectory generation methods typically focus on statistical regularities or spatiotemporal continuity, while overlooking the implicit narrative thread that links activities through motivations, goals, and contextual responses. To our knowledge, few studies have attempted to use narrative as a generative principle for activity sequences, and none have explicitly connected narrative to the dynamic execution of behaviors in simulation. This gap motivates our work, which introduces narrative as both a semantic scaffold for generating coherent activity plans and a bridge to action-oriented simulation.

\section{Methodology}
To move beyond data-centric trajectory generation and toward cognitively grounded behavioral simulation, we propose a Hierarchical LLM-Agent Framework that models human mobility as a layered process of reasoning, planning, and acting. Unlike conventional trajectory synthesis methods that directly output activity coordinates or static schedules, our framework simulates a \textit{micro-society of autonomous agents} whose daily behaviors emerge from a multi-layer cognitive hierarchy.

Conceptually rooted in the \textbf{Belief–Desire–Intention (BDI)} model, the framework integrates large language model (LLM) modules to enable hierarchical reasoning and adaptive reflection. Each agent operates across three interconnected cognitive layers. At the \textbf{macro level}, LLM-based narrative generation formulates personal goals, motivations, and contextual intentions that represent the agent’s desire, and then the narrative is translated into a structured activity plan. At the \textbf{meso level},the reflective planning mechanism dynamically revises decisions based on contextual feedback and occupational flexibility, quantified by the Mobility Entropy by Occupation (MEO) index. At the \textbf{micro level}, the agent grounds its abstract intentions in concrete spatial and temporal actions, selecting destinations and transportation modes within environmental constraints.

This hierarchical organization bridges semantic cognition with embodied mobility, enabling agents to exhibit interpretability, adaptability, and social heterogeneity. The framework’s fidelity is \textbf{validated against real-world mobility diaries} through Jensen–Shannon Divergence–based metrics to ensure consistency between simulated and empirical behavioral patterns.

\subsection{Overall Framework}
\begin{figure}
    \centering
    \includegraphics[width=1\linewidth]{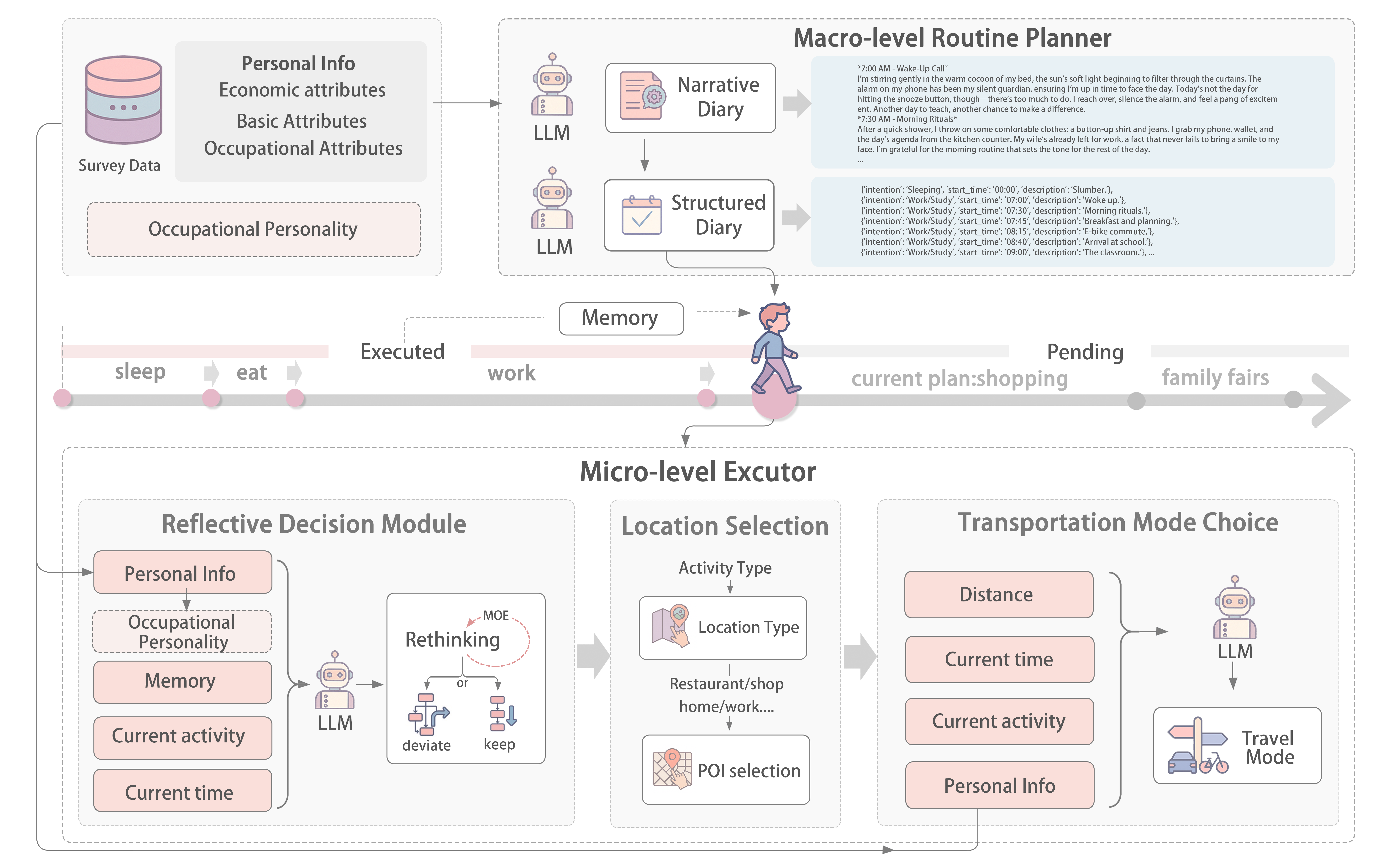}
    \caption{Overall Architecture of the Narrative-to-Action Hierarchical Framework for Human Mobility Simulation}
    \label{fig:framework}
\end{figure}

Figure~\ref{fig:framework} illustrates the architecture of the Hierarchical LLM-Agent Framework, which integrates macro-level narrative reasoning, meso-level reflective planning, and micro-level behavioral execution. Together, these layers implement a cognitively inspired, multi-step reasoning process that mirrors human decision hierarchies. 

At the macro level,  each agent is initialized with a personal profile $p$ derived from  real-world travel surveys. At each time step $t$, the agent is also characterized by a state $s_t$, which encapsulates its current location, memory, and other contextual information. Based on this profile, the narrative generator $\mathcal{N}$ produces a diary-style narrative $z$ containing goals, motivations, and contextual cues. This represents the agent’s \textbf{desire} in the  BDI model. The plan parser $\mathcal{P}$  then translates narrative $z$ into a structured activity plan $A = (a_1, a_2, \dots, a_T)$, forming the agent’s \textbf{intentions} and ensuring both semantic richness and logical coherence:

\begin{align}
z &= \mathcal{N}(p), \label{eq:narrative}\\
A &= \mathcal{P}(z) = (a_1, a_2, \dots, a_T), \label{eq:plan}\\
a_t &= (i_t,\, t_t,\, d_t), \quad t=1,\dots,T. \label{eq:activity}
\end{align}

This design explicitly operationalizes the initial components of the BDI model. An agent’s desires are first captured by the narrative generator $\mathcal{N}$ as a diary-style narrative $z$ rich in motivations and contextual details. The plan parser $\mathcal{P}$ then transforms these narrative desires into a structured activity plan $A$, representing the intentions to which the agent commits for execution. During execution, the agent’s beliefs are represented by its current state $s_{t}$, encompassing its location, memory, time, and contextual information. 

At the meso level, the reflective decision module $\mathcal{R}$ serves as the meta-controller, dynamically evaluating whether to maintain or modify the plan to a new activity $a’_t$ based on the agent’s current \textbf{belief state  ($s_t$)} and the occupation-based flexibility index \textbf{MEO}. This mechanism enables adaptive behavior consistent with hierarchical cognition—where high-level goals are reconsidered under contextual feedback.
\begin{equation}
a'_t = \mathcal{R}(s_t, a_t).
\label{eq:decision}
\end{equation}

At the \textbf{micro level}, two specialized modules ground the agent’s actions in space and time. The location selector $\mathcal{L}$  determines the  spatial location $l_t$  for the next activity $a'_t$  based on spatial constraints and semantic relevance:
\begin{equation}
l_t = \mathcal{L}(a'_t).
\label{eq:location}
\end{equation}

Whenever spatial movement is required, the transportation module $\mathcal{T}$ then selects the most appropriate travel mode $m_t$ given the previously visited location $l_{t-1}$ and the newly chosen location $l_t$, conditioned on the intended activity $a'_t$, the agent’s profile $p$, and the agent’s memory$h_t$:
\begin{equation}
m_t = \mathcal{T}(l_{t-1}, l_t, a'_t, p, s_t).
\label{eq:transport}
\end{equation}

Finally, the executed activities, visited locations, and chosen travel modes across all time steps are combined to form the complete synthetic trajectory:
\begin{equation}
\tau = \{(a'_t, l_t, m_t)\}_{t=1}^T.
\end{equation}
This trajectory $\tau$ serves as the emergent outcome of the narrative-to-action framework, capturing how agents transform high-level narrative plans (desires and intentions) into adaptive mobility behaviors (grounded actions) in space and time.

\begin{figure}
    \centering
    \includegraphics[width=0.5\linewidth]{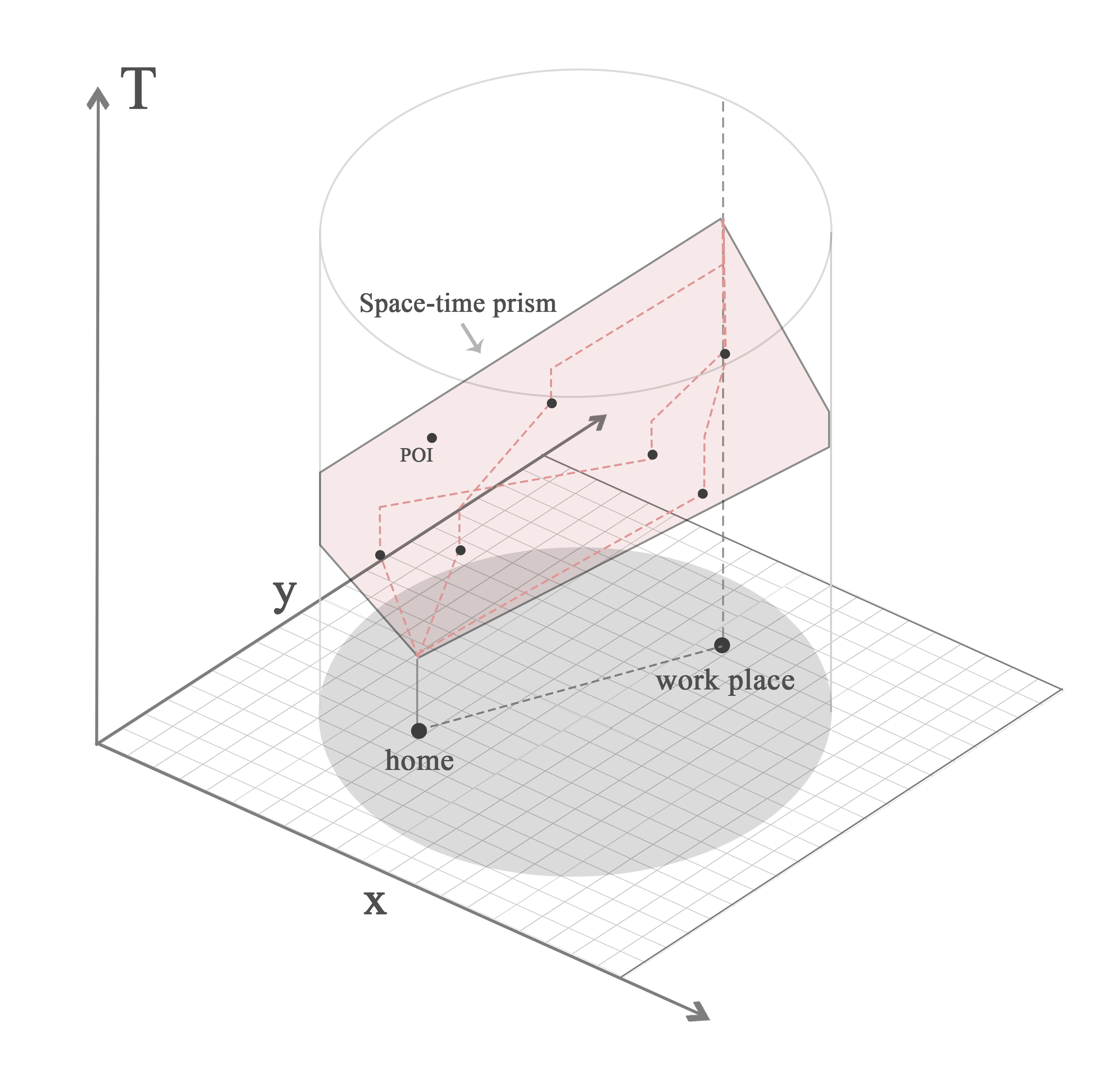}
    \caption{The conceptual space–time framework that governs agent mobility in our simulation. The prism delineates the agent’s possibility space, encompassing all reachable POIs between fixed anchor activities such as home and work.}
    \label{fig:prism}
\end{figure}

Based on time geography theory, all agent activities and movements within this framework are governed by fundamental spatio-temporal constraints. As conceptualized in Figure~\ref{fig:prism}, each agent operates within a 'space-time prism,' which defines the entire set of reachable locations between fixed activities (e.g., being at home in the morning and at work in the afternoon). The activities all occur within these physically plausible boundaries.

\subsection{Macro-level Routine Planner}

The macro-level routine planner is responsible for generating an agent’s initial daily schedule based on its socio-economic profile. To ensure both semantic richness and structural validity, we adopt a two-stage narrative–parsing process. 

\subsubsection{Narrative generation}
In the first stage, we configure the LLM as a “creative writer” tasked with producing a diary-style narrative from a first-person perspective(~\ref{appendix:narrative}). Rather than outputting a rigid timetable, the model generates a coherent, context-rich story that spans the entire day, filled with motivations, emotions, and situational details (e.g., “I felt unusually tired when I woke up, so I grabbed a quick breakfast before rushing to catch the morning bus”). By positioning the LLM in this narrative role, the system ensures that the output resembles a human-like diary, embedding causal links and contextual richness. Formally, this process corresponds to Eq.~\ref{eq:narrative}, where the narrative generator $\mathcal{N}$ produces a diary $z$ given a profile $p$.

\subsubsection{Parsing into structured plans}
In the second stage, we configure another LLM as an “expert in data extraction” whose role is to transform the diary narrative into a structured activity plan(~\ref{appendix:parsing}). The model outputs a machine-readable format (JSON) with explicit attributes such as activity type, start time, and semantic location categories. Unlike the creative writer in Stage 1, this LLM emphasizes precision and consistency, ensuring that the semantic cues embedded in the narrative are faithfully extracted and converted into fields usable for simulation. This process corresponds to Eq.~\ref{eq:plan}–\ref{eq:activity}, where the plan parser $\mathcal{P}$ converts the narrative $z$ into a sequence of activities $A=(a_1,\dots,a_T)$ with explicit attributes $(i_t, t_t, d_t)$.

\subsection{Meso-level reflective decision and Micro-level Executor}

Real human behavior does not always strictly follow planned schedules but tends to involve a certain probability of adjustment. While the macro-level planner generates initial schedules, real human behavior requires adaptive adjustments. The execution stage captures this adaptability through reflective decision-making, location selection, and transportation mode choice.

\subsubsection{Reflective Decision Module with MEO}

The reflective decision module $\mathcal{R}$ is triggered whenever an activity node is completed. At this point, the agent’s state $s_t$ is updated to capture the current time, the most recently executed task, the next planned activity, and its internal memory. Rather than mechanically transitioning to the next scheduled action $a_t$, the decision is delegated to a large language model (LLM). Given the pair $(s_t, a_t)$, the LLM evaluates contextual factors—such as timing, recent events, and the agent’s internal objectives—and determines whether the agent should proceed with the original plan or adapt to an alternative activity (see Appendix~\ref{appendix:rethinking}). This adaptive decision-making process is formally defined in Eq.~\ref{eq:decision}.

To further enhance behavioral realism, we introduce an occupation-aware rethinking mechanism that captures heterogeneity in daily schedule flexibility across different socio-economic groups. Specifically, the probability of triggering a rethinking decision is parameterized by occupation using the MEO:

\begin{equation}
P_\mathrm{rethinking}^{(o)} = MEO_o,
\end{equation}

where $MEO_o$ denotes an occupation-specific constant representing the typical likelihood of deviating from planned activities. For instance, agents representing business owner, who typically have more flexible schedules, are assigned higher values (e.g., $MEO_\mathrm{freelancer} = 0.7$), whereas agents representing factory workers, whose routines tend to be more rigid, are assigned lower values (e.g., $MEO_\mathrm{factory} = 0.3$).

By explicitly modeling these occupational differences, the adaptive execution stage becomes more realistic and socially grounded, enabling the simulation to capture heterogeneous patterns of behavioral flexibility and produce richer, more human-like activity dynamics.

\subsubsection{Location selection}

Once the executed activity $a'_t$ has been determined, the next step is to ground its semantic description into a concrete geographic location. This is handled by the location selector $\mathcal{L}$. For activities such as \emph{working}, \emph{shopping}, or \emph{eating out}, the plan only specifies a semantic category rather than an exact destination. The module $\mathcal{L}$ resolves this abstraction by sampling from the corresponding category of points of interest (POIs) in the simulated city, taking into account contextual information encoded in the agent’s state $s_t$. Formally, this process is described in Eq.~\ref{eq:location}, where $\mathcal{L}$ assigns a specific location $l_t$ to the activity $a'_t$.  
To operationalize this mapping from semantic activities to concrete POIs, we employ a probabilistic destination choice function. This ensures that agents’ spatial movements in the urban simulator are consistent with realistic behavioral patterns. Specifically, we adopt the gravity model, a classic framework in urban science and transportation research(\cite{pappalardo2015returners}). This approach has also been used in the AgentSociety framework(\cite{piao2025agentsociety}). In our formulation, the probability $P_{ij}$ that an agent located at $i$ selects a candidate POI $j$ is defined as:

\begin{equation}
P_{ij} = \frac{D_{ij}^{\beta} A_{j}^{\alpha}}{\sum_{k \in C} D_{ik}^{\beta} A_{k}^{\alpha}}
\label{eq:gravity_model}
\end{equation}

where:$A_j$ is the \textit{attractiveness} of POI $j$, quantified by metrics such as ratings or popularity;$D_{ij}$ is the \textit{travel cost} from $i$ to $j$, typically measured by distance or travel time;$C$ is the set of candidate POIs matching the activity type;$\alpha$ and $\beta$ are tunable parameters controlling sensitivity to attractiveness and travel cost, respectively.

\subsubsection{Transportation mode choice}

The final component of micro-level execution concerns how agents move between activity locations. This module is only triggered when two consecutive activities require a spatial transition, i.e., when the selected locations $l_{t-1}$ and $l_t$ differ. If activities occur in the same location (e.g., transitioning from a family activity to sleeping at home), the transportation step is skipped.

When invoked, the transportation module $\mathcal{T}$ determines the travel mode $m_t$ for the trip from $l_{t-1}$ to $l_t$. The decision considers the agent’s state $s_t$, which incorporates contextual variables such as travel distance and previous activity state, together with the agent’s personal profile $p$ that encodes mobility resources (e.g., car ownership, availability of an e-bike). Instead of using a fixed rule-based system, we delegate the decision-making to an LLM, which selects the most plausible mode given these conditions(~\ref{appendix:transportation}). Formally, this process corresponds to Eq.~\ref{eq:transport}.

\subsection{Evaluation Metrics}
To compare the generated synthetic trajectories with the ground truth dataset, we adopt a set of metrics that capture the essential characteristics of human mobility patterns from two complementary perspectives: spatial and temporal--semantic. As the core statistical tool, we employ the Jensen--Shannon Divergence (JSD) for its effectiveness in measuring the similarity between two probability distributions, as well as its symmetry and smoothness.

JSD is a symmetric variant of the Kullback--Leibler (KL) divergence. Given two probability distributions $P$ and $Q$, it is computed as:
\begin{equation}
\mathrm{JSD}(P \parallel Q) = \frac{1}{2} D_{\mathrm{KL}}(P \parallel M) + \frac{1}{2} D_{\mathrm{KL}}(Q \parallel M)
\end{equation}
where $M = \frac{1}{2}(P + Q)$ is the average distribution, and $D_{\mathrm{KL}}$ denotes the KL divergence. The value of JSD lies in $[0, 1]$, with smaller values indicating greater similarity between the two distributions, meaning the generated trajectories more closely match the feature distributions of the real trajectories.

To compute the four JSD-based evaluation metrics, we first extract feature distributions from both the generated synthetic trajectories and the ground truth dataset. The probability distributions are estimated using normalized histograms, with a small constant $\epsilon$ added to avoid zero probabilities.  

\subsubsection*{(1) Radius of Gyration ($r_g$) --- Spatial dimension} 
Measures the spatial extent of an individual's daily mobility pattern. For a trajectory with $N$ visited locations $\mathbf{r}_i$ and geometric center $\mathbf{r}_{\mathrm{cm}}$, it is computed as:  
\begin{equation}
r_g = \sqrt{\frac{1}{N} \sum_{i=1}^{N} \left\| \mathbf{r}_i - \mathbf{r}_{\mathrm{cm}} \right\|^2 }
\label{eq:radius_gyration}
\end{equation}
The resulting $r_g$ values for all agents are compared between generated and real data using 1D JSD.

\subsubsection*{(2) Daily Unique Locations ($L_d$) --- Spatio-temporal complexity}
Counts the number of distinct POI locations visited by an agent in one day:  
\begin{equation}
L_d = |\{\text{unique POIs visited in a day}\}|
\label{eq:daily_locations}
\end{equation}
The distribution of $L_d$ is compared via 1D JSD to assess diversity and compactness in daily schedules.

\subsubsection*{(3) Transportation Mode Choice --- Modal distribution similarity}  
To evaluate whether generated trajectories reproduce realistic mobility behavior, we examine the distribution of transportation modes chosen for inter-activity trips.  
For each executed trip, we record its transportation mode $m \in \{\text{walk, bike, e\text{-}bike, car, bus, subway}\}$.  
Aggregating across all trips, the empirical probability distribution of modes is computed for both generated and real trajectories:  
\begin{equation}
\mathbf{p}^{(g)} = (p^{(g)}_{\text{walk}}, p^{(g)}_{\text{bike}}, \ldots), \quad
\mathbf{p}^{(r)} = (p^{(r)}_{\text{walk}}, p^{(r)}_{\text{bike}}, \ldots)
\label{eq:mode_distribution}
\end{equation}
The similarity of mode distributions is quantified using JSD:  
\begin{equation}
\mathrm{JSD}_{\mathrm{mode}} = \mathrm{JSD}(\mathbf{p}^{(g)}, \mathbf{p}^{(r)})
\label{eq:transport_mode_jsd}
\end{equation}
This metric captures whether the generated agents select transportation modes in proportions consistent with real-world mobility patterns.

\subsubsection*{(4) Intention Sequences ($s_{\mathrm{intent}}$) --- Temporal ordering patterns}  
Represents the ordered sequence of daily intentions (e.g., \emph{sleep} $\rightarrow$ \emph{work/study} $\rightarrow$ \emph{shopping} $\rightarrow$ \emph{leisure}).  
For each agent, the generated and real trajectories are converted into categorical sequences of intentions aligned by time slots.  
The empirical distributions of these sequences are then compared using 2D JSD, where flattened sequence vectors serve as the basis for probability estimation.  
This metric evaluates whether the generated trajectories reproduce realistic temporal ordering of human activities, beyond aggregate proportions.

\subsubsection*{Final Score}
The four JSD values are aggregated into an overall fidelity score:
\begin{equation}
\text{Final Score} = \frac{1}{4}\sum_{i=1}^{4} (1 - \text{JSD}_i)
\label{eq:final_score}
\end{equation}
where $\text{JSD}_i$ denotes the divergence value of the $i$-th evaluation metric 
(\textit{Intention Seq.}, \textit{Visited Loc.}, \textit{Travel Mode}, and \textit{Turning Radius}). 
Each term $(1 - \text{JSD}_i)$ converts a divergence value into a similarity score, 
and the arithmetic mean over the four metrics yields the Final Score. 
Higher values of the Final Score therefore indicate greater overall fidelity of the generated trajectories to the real data.

\subsection{Dataset and simulation environment}

The validation data for this study are derived from a comprehensive travel diary survey conducted in Guangzhou, comprising approximately 2000 individual activity records collected between November 2018 and January 2019. 
The survey covered seven residential communities in the Tangxia area of Tianhe District with diverse built environments, including commodity housing estates, urban villages, and public rental housing, over a total area of about 2.2\,km\(^2\).

Each participant contributed multi-source data: (1) a household and personal questionnaire capturing demographics, housing status, employment, relocation history, health and lifestyle; (2) five consecutive days of detailed travel diaries documenting activity chains, motivations, visited locations, and transportation modes for both workdays and rest days; (3) GPS trajectories continuously logged during the same period; (4) environmental exposure data such as real-time air quality and noise levels; (5) ecological momentary assessment (EMA) surveys collected four times per day (8:00, 12:00, 16:00, 20:00) on affect and perceptions; and (6) crowd density sensing through \textit{BatSound Quadrant} devices, which measure the number of mobile terminals and signal attenuation within a 50\,m radius for each diary segment.

\begin{figure}
    \centering
    \includegraphics[width=1\linewidth]{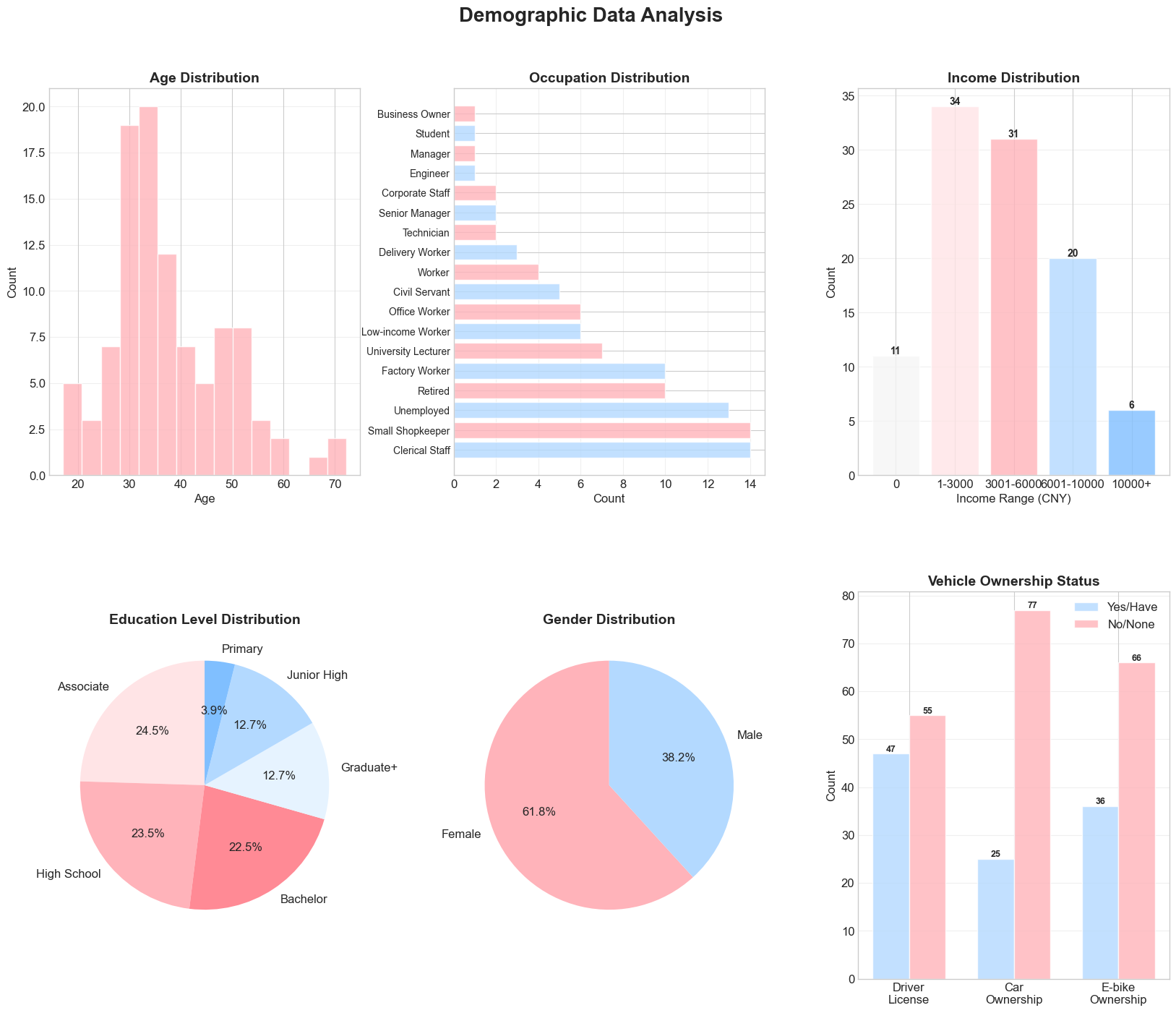}
    \caption{Demographic and socio-economic distributions of the surveyed population, including age, occupation, income, education level, gender, and vehicle ownership.}
    \label{fig:datset}
\end{figure}

The dataset (Fig.~\ref{fig:datset}) thus contains two main components: socio-economic attributes of participants — including demographic information and inferred occupational categories — and daily travel diaries that serve as the empirical benchmark for evaluation. 
These socio-economic attributes are used to construct the personal profiles of agents, while the diary data provide rich ground truth for validating the generated mobility patterns.

To support the occupation-aware rethinking mechanism, we derived occupation categories from the socio-economic attributes and self-reported job descriptions in the survey. 
Based on domain knowledge and observed flexibility patterns in the activity diaries, each occupation group was assigned a fixed MEO
value representing its typical probability of plan adaptation. 
Table~\ref{tab:meo} summarizes the mapping used to parameterize rethinking probabilities in the simulation.

\begin{table}[htbp]
\centering
\caption{Occupation-specific MEO values used to parameterize rethinking probability.}
\label{tab:meo}
\begin{tabularx}{\linewidth}{Xc} 
\toprule
\textbf{Occupation Category} & \textbf{MEO value} \\
\midrule
Factory Worker, Clerical Staff, Delivery Worker, Technician, Low-income Worker & 0.30 \\
University Lecturer, Civil Servant, Engineer, Office Worker, Student & 0.50 \\
Small Shopkeeper, Business Owner, Manager, Senior Manager, Corporate Staff & 0.70 \\
Unemployed, Retired & 0.20 \\
\bottomrule
\end{tabularx}
\end{table}

To ensure data quality and suitability for modeling, several pre-processing steps are applied. All survey records are anonymized to protect privacy. Incomplete or inconsistent trajectories are removed, and outliers (e.g., unrealistically long travel times) are filtered. Time information is discretized into 15-minute intervals for consistency with the simulation granularity. Free-text diary descriptions are parsed into structured activity categories (e.g., work, shopping, leisure), and missing attributes are inferred where possible. These procedures produce a clean, structured dataset that serves as both the input for agent initialization and the benchmark for evaluation.

To construct the urban environment, we supplement the survey data with spatial information from OpenStreetMap (OSM), including road networks, area-of-interest (AOI) polygons, and POI categories. The integration of survey-based behavioral data and OSM-based environmental data enables both realistic profile initialization and geographic grounding of activities. All simulations are conducted on the AgentSociety platform \cite{piao2025agentsociety}, which provides a scalable infrastructure for large-scale, LLM-driven agent-based modeling.

\subsection{Implementation Details}
\textbf{Model configuration.}  
We employ the GLM-4 Flash model(\cite{glm2024chatglm}) for all LLM components. 
In early experiments, we evaluated more advanced and larger-scale models, but none outperformed GLM-4 Flash. 
This observation aligns with recent findings that bigger models are not necessarily superior for human-scale behavioral tasks(\cite{wilcox2025bigger}). 
Given the need to control inference cost and to support high concurrency in large-scale simulations, we selected GLM-4 Flash as our unified backbone. 
The sampling temperature is set to 1 to encourage diversity in generated activity plans.

\section{Results}
\subsection{Quantitative Results}

\begin{table*}[t]
\centering
\caption{Ablation study of the proposed framework. Lower JSD values indicate better alignment in each metric; higher \emph{Final Score} indicates better overall fidelity. \textbf{Bold} numbers denote the best within each block; \underline{underlined} numbers denote the second best.}

\resizebox{\textwidth}{!}{%
\renewcommand{\arraystretch}{1.5}
\begin{tabular}{l|ccc|cc|cc|ccccc}
\toprule
\multirow{2}{*}{\textbf{System Configuration}} 
& \multicolumn{3}{c|}{\textbf{Plan Generation}} 
& \multicolumn{2}{c|}{\textbf{Travel Mode Choice}} 
& \multicolumn{2}{c|}{\textbf{Rethinking}} 
& \multicolumn{5}{c}{\textbf{Evaluation Metrics}} \\
\cmidrule(lr){2-4} \cmidrule(lr){5-6} \cmidrule(lr){7-8} \cmidrule(lr){9-13}
& Random & LLM-Direct & Narrative$\to$Parsing 
& Random & LLM 
& None & Prob. 
& Intention Seq. $\downarrow$ & Visited Loc. $\downarrow$ & Travel Mode $\downarrow$ & Turning Radius $\downarrow$ & Final Score $\uparrow$ \\
\midrule

Full Model w/ Random Plan       & \checkmark &  &  &  & \checkmark &   & \checkmark  & 0.531 & 0.698 & 0.212 & 0.349 & 0.552 \\
Full Model w/ Direct LLM Plan   &  & \checkmark &  &  & \checkmark &   & \checkmark  & 0.466 & 0.441 & 0.159 & \underline{0.361} & 0.6434 \\
Full Model w/ Random choice     &  &  & \checkmark & \checkmark &  &  & \checkmark & \underline{0.460} & 0.486 & 0.254 & \textbf{0.345} & 0.614 \\
Full Model w/o Rethinking       &  &  & \checkmark &  & \checkmark & \checkmark &  & 0.475 & \textbf{0.250} & \underline{0.138} & 0.466 & \underline{0.668} \\
Full Model w/ Prob. Rethinking  &  &  & \checkmark &  & \checkmark &  & \checkmark & \textbf{0.431} & \underline{0.386} & \textbf{0.126} & 0.369 & \textbf{0.672} \\
\bottomrule
\end{tabular}%
}
\label{tab:ablation_final}
\end{table*}
The quantitative results of the ablation study Table~\ref{tab:ablation_final} systematically evaluate the contribution of each component within the proposed framework, confirming that the full model configuration achieves the highest overall fidelity with a Final Score of 0.672. The data first demonstrates the superiority of the two-stage narrative-to-parsing approach, which significantly outperforms both direct LLM generation (Final Score: 0.6434) and random planning (Final Score: 0.552) by producing more realistic activity sequences and travel mode choices. Furthermore, the study highlights the critical role of an intelligent travel mode selection module; replacing the LLM with a random choice causes the Travel Mode JSD to nearly double from 0.126 to 0.254, underscoring the necessity of a context-aware mechanism. Finally, the inclusion of a probabilistic rethinking mechanism provides the decisive enhancement, pushing the model to its peak performance by improving key behavioral metrics such as Intention Sequence and Travel Mode. Therefore, the study validates that the framework's success is contingent upon the synergy of semantically rich planning, intelligent execution, and dynamic behavioral adaptation.

 \subsection{Qualitative Analysis}
To illustrate the diversity and interpretability of the generated diaries, we highlight two representative cases that reflect distinct daily mobility patterns (see~\ref{appendix:sample-diaries}).The sampleA depicts a home-oriented weekday in which the individual structures the day around domestic and personal activities. The morning is devoted to coffee, checking notifications, and applying for jobs, followed by household cleaning and gardening. Lunch is simple and prepared at home, and the afternoon includes light chores, newspaper reading, and rest. Apart from a short evening stroll around the neighborhood, mobility is absent, making walking the sole travel mode. Spatially, the activity locations are concentrated within the dwelling and its immediate surroundings, and social interactions are limited to indirect information exchanges such as online news or community notices. Overall, this pattern exemplifies a low-mobility, home-centered lifestyle where activities are highly clustered and interaction intensity is minimal.

The sample B presents a contrasting commuting workday characteristic of a university lecturer. The day begins with a morning routine and a car-based commute to campus, followed by tightly scheduled teaching sessions, departmental meetings, and administrative work. Within the campus, short walking trips connect the office, lecture halls, and cafeteria, while the evening includes a return commute and a neighborhood dog walk. This profile illustrates a bimodal commuting structure, higher spatial dispersion across residential, workplace, and campus facilities, and more intensive social engagement through meetings, classroom interactions, and informal conversations with colleagues. Compared to the home-based case, this routine is more structured, spatially diverse, and socially embedded.

Taken together, these two examples demonstrate that the generated narratives can capture meaningful heterogeneity in daily life. The model reproduces differences in activity composition (domestic tasks versus professional obligations) and interaction intensity (minimal versus frequent). This suggests that the framework is capable of generating realistic and distinguishable daily patterns aligned with empirical observations of both non-working, home-centered individuals and actively employed, commuting populations.

\section{Discussion}

The Narrative-to-Action hierarchical framework proposed in this study has achieved encouraging results in generating high-fidelity human mobility trajectories. By decomposing mobility generation into macro-level narrative planning, meso-level reflective adjustment, and micro-level behavioral execution, the framework achieves a balance between interpretability and adaptability rarely observed in prior data-driven approaches. Unlike traditional sequence models that treat mobility as a static spatio-temporal prediction problem, this hierarchical design allows emergent behaviors to arise from multi-level cognitive reasoning processes.

This hierarchical organization enables each component to specialize in distinct cognitive functions—reasoning, intention formation, and action execution—while maintaining an adaptive feedback loop among layers. In essence, the Narrative-to-Action framework operationalizes a \textit{cognitively hierarchical LLM-Agent paradigm}, bridging narrative planning with embodied simulation. We attribute the framework’s success to the complementary roles of narrative coherence and the reflective rethinking mechanism. First, the two-stage process of narrative generation and parsing ensures that an agent's initial activity plan is not only structurally valid but also semantically rich and internally logical. Unlike methods that directly generate sequences of spatio-temporal coordinates, the narrative endows each activity with clear motivation and context. This context-rich intention provides a solid foundation for subsequent, more realistic behavioral decisions, particularly in transportation mode choice. Second, the reflective mechanism in the dynamic execution stage allows agents to flexibly adjust their plans according to their current state, breaking the rigidity of static schedules and thereby enhancing behavioral realism.

The introduction of the MEO further enriches this hierarchy by encoding social heterogeneity into adaptive decision-making. Through MEO, occupation-specific flexibility becomes an intrinsic property of agents, leading to more diverse and socially grounded mobility outcomes. MEO is more than a technical parameter; it serves as a bridge between micro-level behavioral simulation and macro-level socio-economic theory. Compared to works like TrajLLM(\cite{ju2025trajllm}) and MobAgent(\cite{li2024more}), which primarily focus on generating static activity plans, our framework, through its MEO-driven dynamic execution, is the first to simulate the behavioral flexibility of individuals during plan execution, arising from an individual's professional characteristics and social roles.

The proposed framework offers valuable tools for urban transportation planning and policy evaluation. Urban administrators can use the privacy-preserving, high-fidelity synthetic data generated by our framework for travel demand forecasting, infrastructure planning, and traffic management optimization.  Beyond trajectory realism, the framework reveals new possibilities for explainable, policy-relevant mobility modeling—each decision can be traced to a cognitive rationale expressed in natural language.  For example, the narrative layer can be extended to simulate behavioral responses to interventions such as flexible work policies or transit disruptions, while the reflective controller could support counterfactual experimentation at the agent level. Furthermore, the MEO-driven adaptive agents can be used to assess the differentiated responses of various social groups to policy interventions. For instance, when evaluating a new congestion pricing policy, the model could predict that business professionals with high MEO (flexible schedules) might choose to travel during off-peak hours or work remotely, whereas factory workers with low MEO (rigid schedules) might be forced to bear higher travel costs or switch to more time-consuming public transport. This ability to capture socially heterogeneous responses opens up opportunities for studying equity-oriented interventions.

Despite the framework's advantages, we acknowledge several limitations that also point to future research directions. First, the dataset used for constructing personal profiles and for validation is from a specific area and is relatively limited in scale, which may affect the generalizability of the results. Future work should focus on validating and extending this framework with more diverse datasets from different cultural backgrounds and city sizes. Second, the agent interaction mechanisms in the current model are simplified and do not fully capture complex social network dependencies. Future research could incorporate social network structures, such as family and colleague relationships, to simulate more realistic group co-decision-making. Finally, considering the computational demands of large-scale urban simulations, we deliberately chose the efficient GLM-4 Flash model. An important future direction is to explore the use of smaller, locally deployable, and specialized language models to further enhance simulation scalability and reduce costs while maintaining behavioral realism.

\section{Conclusion}

This study presents a Narrative-to-Action \textbf{Hierarchical LLM-Agent Framework} that bridges narrative reasoning and adaptive behavioral simulation, providing a cognitively interpretable approach to modeling human mobility. By explicitly structuring the agent’s cognition into \textbf{macro-level narrative generation}, \textbf{meso-level reflective planning}, and \textbf{micro-level action execution}, the proposed framework advances beyond conventional flat generation models. By decoupling semantic plan generation from dynamic action selection, our framework overcomes the rigidity of static schedules, producing daily diaries that are both interpretable and behaviorally realistic. The incorporation of the \textbf{MEO-driven reflective mechanism} further enhances the system’s ability to simulate heterogeneous behavioral flexibility across occupational groups.

Empirical evaluations demonstrate that this hierarchical design significantly improves the fidelity and interpretability of synthetic mobility data. Beyond generating privacy-preserving trajectories, the framework offers new opportunities to understand \textit{how} human decision processes unfold hierarchically—from goals and intentions to adaptive actions—in complex urban environments. In essence, this work redefines mobility simulation as a hierarchical cognitive process, positioning LLM agents not merely as text generators but as multi-level reasoning entities capable of bridging semantics, cognition, and behavior. Future extensions could incorporate social interaction layers and localized language models, further evolving toward city-scale, socially grounded cognitive simulations.

\section*{Declaration of generative AI and AI-assisted technologies in the writing process}

During the preparation of this work the author(s) used ChatGPT  in order to improve the readability and clarity of the manuscript. After using this tool, the author(s) reviewed and edited the content as needed and take(s) full responsibility for the content of the published article.

\section*{Data Availability Statement}
The data that support the findings of this study are available from the corresponding author upon reasonable request.

\bibliographystyle{elsarticle-harv}
\bibliography{refs} 

\appendix
\section{Prompt Design for Narrative-to-Action Framework}

\subsection{Narrative generation}
This prompt instructs the large language model to generate a first-person daily activity narrative based on a specified personal profile.
\label{appendix:narrative}

\subsubsection{System Prompt}
\begin{verbatim}
You are a citizen. Write a realistic, first-person daily log about your own day, 
from waking up to going to sleep. Be descriptive and include approximate times 
for each activity.
\end{verbatim}

\subsubsection{User Prompt}
\begin{verbatim}
{character_profile}

**Your Task:**
Write a plausible, chronologically ordered story of this character's day. 
Describe their activities, what they might be thinking, and roughly when they 
do things. Make it sound like a real person's day.

**Example for a Programmer with a car:**
"I woke up around 8:00 AM, scrolled through my phone for a bit before getting up..."

**Example for a student without a car:**
"I woke up around 7:30 AM, got ready quickly, and took the subway to university..."

**Example for an unemployed person:**
"I woke up around 8:30 AM, had a leisurely breakfast, and spent the morning..."

**Now, please generate the narrative for the character described above, 
considering their transportation options and work status.**
\end{verbatim}

\subsection{Parsing into structured plans}
This prompt is used to parse and transform the natural language narrative generated in the previous step into a structured JSON plan.
\label{appendix:parsing}

\subsubsection{System Prompt}
\begin{verbatim}
You are an expert data extraction and classification tool. Your task is to 
read a daily log and convert it into a structured JSON plan. You must classify 
each activity into a predefined category.
\end{verbatim}

\subsubsection{User Prompt}
\begin{verbatim}
**Activity Categories:**
{activity_categories}

**Source Narrative:**
---
{narrative}
---

**Your Task:**
Analyze the narrative above and extract a chronologically sorted list of activities. 
For each activity, determine its start time and classify it into one of the 
categories provided.

**Output Format (Strictly JSON):**
- Respond with a single JSON object with a key "plan".
- The `activity` MUST be one of the specified activity categories.
- The `start_time` MUST be in “HH:MM” format (24-hour clock).
- The `description` should be a brief summary from the narrative.
- The first activity should start at "00:00" and be "sleep".

**Example:**
{example_json}
\end{verbatim}

\subsection{Reflective decision-making}
This prompt simulates the agent’s internal decision-making process, determining whether to follow the existing plan or adapt it according to the circumstances.
\label{appendix:rethinking}

\subsubsection{System Prompt}
\begin{verbatim}
You are the inner voice of a person. Based on your current feelings, 
past experiences, and plans, decide if you should do something else instead. 
Be brief and decisive.
\end{verbatim}

\subsubsection{User Prompt}
\begin{verbatim}
{character_profile}

Current time: {formatted_time}
{memory_context}

**Activity Categories:**
{activity_categories}

**IMPORTANT**: You should consider different occupations’ Occupational Personality 
Entropy, which reflects the degree of uncertainty or flexibility in their daily 
plans...

Based on your current status:

**Output Format (Strictly JSON):**
Based on the guidelines, provide your decision in the following JSON format.

**1. If you follow the plan:**
{{
  "action": "follow",
  "reasoning": "A brief reason why you are sticking to the plan."
}}

**2. If you change the plan:**
{{
  "action": "change",
  "new_activity": "A new activity name from the list: {activity_categories}",
  "duration_minutes": <A plausible duration in minutes, e.g., 90>,
  "reasoning": "A brief reason for the change."
}}
\end{verbatim}

\subsection{Transportation mode choice}
This prompt is used to decide the most appropriate travel mode based on detailed trip information such as distance, purpose, and time.
\label{appendix:transportation}
\subsubsection{System Prompt}
\begin{verbatim}
You are the inner voice of a person making a practical, everyday transportation 
decision. Think step-by-step and choose the most sensible option.
\end{verbatim}

\subsubsection{User Prompt}
\begin{verbatim}
You need to decide how to get to your next destination. Here's your situation:
{character_profile}

**The Trip:**
- Destination: '{destination_poi_name}' (a {destination_poi_type})
- Purpose of trip (your intention): {activity_type}
- Distance: Approximately {distance} meters.
- Current Time: {formatted_time}

**Your Task:**
Based on all of this, what's the most logical way for you to travel? Consider 
factors like distance, time, weather (assume it's normal) and cost.

Please respond in a strict JSON format with your choice and a brief reasoning.
You choice answer must be one of them: {available_options}

Example Response:
{{
    "reasoning": "The distance is short... so walking is a good choice...",
    "choice": "Walking"
}}

Another Example:
{{
    "reasoning": "It's quite far... Driving is the fastest...",
    "choice": "Driving"
}}

Now, provide your decision for your current situation.
\end{verbatim}

\section{Sample Diaries}
\label{appendix:sample-diaries}
\subsection{Diary ExampleA}
The day began at 7:30 AM with waking up late, grabbing a cup of steaming coffee, and quietly gazing out the window at the overcast sky. After the first sip of coffee brought a bit of energy, the phone was checked for notifications, followed by jotting down a simple routine that included checking the mail, browsing job listings, and doing some household chores. By 8:30 AM the mail had arrived—mostly bills, flyers, and catalogs—before turning attention back to job listings. Despite frustration with an outdated resume, a few applications were sent out by 9:00 AM, mainly for part-time or remote positions, along with setting a reminder to call the local job service office. At 9:30 AM the focus shifted to cleaning the kitchen, tackling piled-up dishes, and wishing for a helper.  

At 10:30 AM there was a break outside, breathing in the cool air, checking on the roses in the garden, watering them, and enjoying the beauty of nature. Returning indoors at 11:00 AM, a simple sandwich with mustard and lettuce was prepared for lunch. At 11:30 AM the laundry was started, with towels and jeans in the washing machine and fabric softener sheets added for freshness. By 1:00 PM, there was time to read the newspaper, noting community events and adding a few to the calendar. At 2:00 PM the living room was cleaned, furniture dusted and rearranged, though the work left a sense of tiredness. At 3:00 PM a short rest followed with another cup of coffee while watching an old movie, offering a comfortable rhythm to the afternoon.  

As the sun began to set, around 5:00 PM, dinner preparation started, leading to a pasta dish with garlic bread. At 6:30 PM dinner was enjoyed alone, savored with a sense of accomplishment. At 7:30 PM, a quick walk around the block provided fresh air, passing a park where children played, which brought back memories of raising children. Returning home at 8:00 PM, the evening was spent watching a crime drama on television, providing relaxation before bedtime. By 9:30 PM, preparations for sleep were made—brushing teeth, setting the alarm, and reflecting on the day’s job applications with a small glimmer of hope for the future. Finally, at 10:00 PM, the day ended peacefully, with a sense of contentment and optimism for progress in the job search tomorrow.  
\subsection{Diary Example B}

 The day began at 6:15 AM with the gentle buzz of the phone alarm. After hitting snooze twice, a quick shower washed away the last traces of sleep, followed by a simple breakfast of cereal with almond milk and mixed berries. By 7:10 AM, dressed in professional attire—a navy suit and crisp white shirt—the commute began in a reliable car. Traffic was lighter than expected, and by 8:00 AM the office was reached smoothly. The morning started with emails and urgent messages before a sequence of meetings at 8:45 AM, where ongoing projects were discussed and progress was made. A brief stroll at 9:30 AM offered fresh air before returning to mark student papers and prepare the following day’s lecture.  

At noon, lunch was taken in the cafeteria with a colleague, accompanied by light conversation about weekend plans and a documentary series. Back in the office at 1:00 PM, the focus shifted to teaching. The lecture hall filled quickly, and the session successfully engaged students with a challenging topic. After class, a coffee break in the staff room allowed time to chat with fellow faculty before returning to the office at 3:45 PM to complete feedback on assignments. As the workday wound down at 5:15 PM, belongings were packed, and by 5:40 PM the commute home began, ending at 6:20 PM in a quiet apartment where the family dog eagerly waited.  

Dinner at 6:30 PM was a simple pasta with tomato sauce prepared earlier, followed by a relaxing evening walk with the dog at 7:45 PM. The cool evening air and city lights provided a chance to unwind. Returning home at 8:15 PM, some time was spent reading before preparing for bed at 9:00 PM. A final phone check at 9:45 PM concluded the day, closing with a sense of productivity and anticipation for the challenges of tomorrow. This entry reflects a typical Wednesday in the life of a university lecturer, structured around commuting, teaching, and family routines.  

\end{document}